\documentclass[final,12pt,3p,times]{elsarticle}

\usepackage{lineno}

\usepackage{amsmath}
\usepackage{amstext}
\usepackage{amsgen}
\usepackage{amsfonts}
\usepackage{amssymb}

\usepackage{array}
\usepackage{graphicx}
\usepackage{hyperref}
\usepackage[framed,autolinebreaks]{mcode}

\newtheorem{theorem}{Theorem}
\newtheorem{corollary}{Corollary}

\journal{Metrology 2023, 3(2), 222-236; \url{https://doi.org/10.3390/metrology3020012}, }

\begin{document}

\begin{frontmatter}



\title{Characteristic Function of the Tsallis $q$-Gaussian and Its Applications in Measurement and Metrology}


\author{Viktor Witkovsk\'y}

\ead{witkovsky@savba.sk}



\address{Institute of Measurement Science, Slovak Academy of Sciences, Dúbravská Cesta 9, 841 04 Bratislava, Slovakia}

\begin{abstract}
The Tsallis $q$-Gaussian distribution is a powerful generalization of the standard Gaussian distribution and is commonly used in various fields, including non-extensive statistical mechanics, financial markets and image processing. It belongs to the $q$-distribution family, which is characterized by a non-additive entropy. Due to their versatility and practicality, $q$-Gaussians are a natural choice for modeling input quantities in measurement models. This paper presents the characteristic function of a linear combination of independent $q$-Gaussian random variables and proposes a numerical method for its inversion. The proposed technique makes it possible to determine the exact probability distribution of the output quantity in linear measurement models, with the input quantities modeled as independent $q$-Gaussian random variables. It provides an alternative computational procedure to the Monte Carlo method for uncertainty analysis through the propagation of distributions.
\end{abstract}




\end{frontmatter}


\renewcommand{\baselinestretch}{1.2}

\section{Introduction}\label{sec1}

{According to Supplement~1 
 \cite{GUMS1} of the {Guide to the Expression of Uncertainty in Measurement} (GUM) \cite{GUM}, a method for evaluating measurement uncertainty consists of three main stages: formulation, propagation and summarization.} 
The first stage, formulation, is particularly crucial, as it involves a series of important steps that must be taken in order to accurately determine the measurement output quantity or measurand. These steps include defining the measurand, identifying the quantities that influence it, developing a model or measurement equation that relates the output to the inputs and assigning probability density functions (PDFs) to the input quantities based on available knowledge. Typically, this knowledge is derived from direct measurements and expert knowledge. In addition to the common PDFs such as Gaussian (normal) and rectangular (uniform) distributions, other distributions based on reasonable principles and available information may also be~used.

In some cases, it may be necessary to assign PDFs to quantities that have not been explicitly measured or for which only partial information is available. The principle of maximum entropy ( {Entropy is a fundamental concept that finds applications in various scientific and engineering fields such as measurement, probability, statistics and information theory. It was introduced by William Rankine, Rudolf Clausius, Ludwig Boltzmann, Josiah Willard Gibbs, James Clerk Maxwell and other scientists in the second half of the 19th century. Claude Shannon later expanded on the concept of entropy in information theory. Entropy provides a useful tool for characterizing uncertainty, randomness and information content in different systems. The higher the entropy, the more disordered or uncertain the system is. In measuring uncertainty, entropy quantifies the uncertainty in a measurement by characterizing the probability distribution of the measurement result. The principle of maximum entropy states that the probability distribution that best represents the current state of knowledge about a system is the one with the largest entropy. This principle was first articulated by Edwin Thompson Jaynes in 1957}) is
 a valuable tool for this task, as it allows us to construct a PDF that accurately characterizes our incomplete knowledge of a quantity. This involves maximizing the traditional entropy, as defined, e.g., by Shannon, subject to constraints imposed by the available information. The principle of maximum entropy is particularly useful in situations where there are no indications available and we must rely solely on the available information to represent the PDF of a given quantity. To learn more about the principle of maximum entropy and its application in measurement and uncertainty evaluation, {consult the Supplements to the GUM \cite{GUMS1,GUMS2}.}
	
The Tsallis $q$-Gaussian distribution is a probability distribution introduced by Tsallis~\cite{tsallis1988} as a generalization of the standard normal (Gaussian) distribution based on maximizing the  
Tsallis entropy 
  under appropriate constraints. It belongs to a larger family of probability distributions known as $q$-distributions, characterized by a non-additive entropy that generalizes the Boltzmann--Gibbs entropy used in statistical mechanics; for more details, see, e.g., \cite{tsallis2009}. Non-additive entropy refers to a family of entropy measures that do not satisfy the additivity property of the traditional additive entropy measures. Additivity means that the entropy of a joint system can be obtained by adding the entropy of the individual systems. However, in some cases, additivity may not hold, particularly for complex systems with non-linear interactions between their components.

Non-additive entropy measures are used in various fields, such as physics, information theory and economics, to quantify the degree of uncertainty or disorder in a system that involves non-linear interactions. Examples of non-additive entropy measures include Tsallis entropy and Renyi entropy, which are widely used in the study of complex systems and statistical mechanics. These measures have been found to be more suitable for modeling complex systems than the traditional additive entropy measures.

Compared to the standard Gaussian distribution the spread of the $q$-Gaussian distribution depends on both the scale parameter and the $q$-index parameter which leads to different behavior in the tails of the distribution. Depending on the value of the $q$-index parameter, the $q$-Gaussian distribution may have unique properties that make it useful for modeling specific data. When the value of the $q$-index parameter is less than one, the distribution has a finite support, meaning that it is bounded on both sides. This can be useful for modeling data that are known to have a specific range, such as time intervals or distances. On the other hand, when the value of the $q$-index parameter is greater than or equal to $5/3$, the variance of the $q$-Gaussian distribution does not exist, meaning that the distribution has infinite variance. Moreover, when $2 \leq q < 3$, other moments of the distribution may also not exist or may not be well-defined and the $q$-Gaussian distribution is located in a region of extremely heavy tails. This can be useful for modeling data that has extreme values, such as data with outliers or natural phenomena with rare but significant~events.

{The $q$-Gaussian distribution is a versatile and practical tool that finds potential applications in various fields such as physics, astronomy, geology, anatomy, economics and finance, molecular biology and engineering. Its ability to model complex systems with long-range interactions, memory effects or non-equilibrium dynamics makes it a useful distribution for real industry applications and as a modeling distribution of input quantities in measurement models.

In statistical mechanics, $q$-Gaussian distributions can describe the velocity distribution of particles in non-extensive systems such as turbulent fluids, plasmas and granular gases. In geology, $q$-Gaussian distributions can fit the frequency-magnitude distribution of earthquakes and volcanic eruptions. In anatomy, $q$-Gaussian distributions can model the distribution of human inter-beat intervals and brain activity. In astronomy, $q$-Gaussian distributions can represent the distribution of velocities of stars in globular clusters and galaxies. In economics, $q$-Gaussian distributions can capture the fat-tailed behavior of financial returns and volatility. In machine learning, $q$-Gaussian distributions can be used as activation functions or basis functions for neural networks and radial basis\linebreak function~networks.

Vignat and Plastino \cite{vignat2009} explore the possible reasons why $q$-Gaussian distributions are frequently observed in various natural and artificial phenomena. They argue that the detection of $q$-Gaussian behavior may be influenced by the normalization process performed by the measurement device. If the incoming data have elliptical symmetry, a common property of many distributions, then the normalized data will always follow a $q$-Gaussian distribution, with a parameter $q$ that depends on the normalization technique.

Beck et al. \cite{beck2005} show that the velocity distribution of particles in a turbulent fluid can be modeled via a $q$-Gaussian distribution, with a parameter $q$ that depends on the Reynolds number, and propose a superstatistical framework to explain the origin of $q$-Gaussian behavior in complex systems. Several studies, e.g., \cite{borland2005}, have found $q$-Gaussian behavior in the returns and volatility of financial assets, with different values of $q$ for different markets and time scales. Carpena et al. \cite{carpena2009} analyzed the distribution of distances between consecutive occurrences of a given nucleotide in DNA sequences and found that it follows a $q$-Gaussian distribution with $q$ close to 2. Burlaga et al. \cite{burlaga2004} studied the distribution of fluctuations in the magnetic field and plasma density of the solar wind and observed that it is well-fitted by a $q$-Gaussian distribution with $q$ around 1.6 and relates the value of $q$ to the nonextensive entropy parameter. Anteneodo in \cite{anteneodo2005} investigates the nature of random variables whose sums lead to $q$-Gaussian distributions and proposes a simple statistical mechanism based on non-extensive random walks. The paper also discusses some examples of applications of $q$-Gaussian distributions and non-extensive random walks in physics, biology and finance.

In measurement, it is common to encounter the challenge of having only a small number of measurement repetitions. When using the Type A evaluation method with $n\leq 3$ measurements that have Gaussian errors, we need to use the Student's $t$ distribution with $\nu\leq 2$ degrees of freedom to model the distribution of the input variable. However, in this case the variance of the distribution does not exist, which requires caution when evaluating the associated measurement uncertainty. Nonetheless, as will be explained later, this distribution can also be represented as a $q$-Gaussian distribution with $q \geq 5/3$.
}

Combining independent $q$-Gaussian random variables is a powerful tool for modeling the behavior of a measurand that involves diverse and complex input quantities. The\linebreak $q$-Gaussians offer a flexible family of distributions that can capture a wide range of statistical behaviors, making it a suitable choice for such modeling tasks.

In this paper, we propose an approach to assess the probability distribution of output quantities in linear measurement models using the {Characteristic Function Approach} (CFA), as described in \cite{witkovsky2017}. An alternative approach is through the propagation of distributions using Monte Carlo methods, as outlined in the Supplements of the GUM \cite{GUMS1, GUMS2}.

The paper is arranged as follows. Section~\ref{sec2} presents an overview of the characteristic function approach for assessing the measurement uncertainty. In Section~\ref{sec3}, we derive  the characteristic function of the $q$-Gaussian random variables and present a numerical method for inverting the characteristic function (CF) of a linear combination of independent $q$-Gaussian random variables, say $Y = \sum_{k=1}^n c_k X_k$.  Section~\ref{sec4} describes the basic functionality of the MATLAB toolbox \href{https://github.com/witkovsky/CharFunTool}{\textbf{CharFunTool}} which
 provides a collection of characteristic functions of various probability distributions and algorithms for their manipulation and inversion.  Section~\ref{sec5} presents some illustrative numerical examples. Section~\ref{sec6} includes discussion and conclusions.
\section{Characteristic Function Approach for Assessing the Measurement Uncertainty}
\label{sec2}
{There are various methods and approaches for assessing the measurement uncertainty of a quantity that depends on multiple input quantities characterized with different probability distributions. The most commonly used methods are those advocated in \cite{GUM, GUMS1, GUMS2}. However, in certain cases, when dealing with distributions such as the Tsallis $q$-Gaussian distribution with a $q$-index greater than or equal to $5/3$, the standard uncertainty of some of the considered input quantities may not exist. In such cases, the approach to evaluating the uncertainty using the law of propagation of uncertainty, as specified in \cite{GUM}, may not be applicable. Nonetheless, alternative approaches can be used to assess the uncertainty by evaluating the coverage intervals in these situations and the expanded uncertainty can be formally stated as the half-width of the interval.

Here we adopt the approach of GUM Supplements \cite{GUMS1, GUMS2} which consider the output distribution as a posterior distribution for the measurand obtained by propagating the distributions of the input quantities through the measurement model. This posterior distribution is then utilized to calculate an interval or region that encompasses a large fraction of the distribution of values that could reasonably be attributed to the measured quantity based on available information. Based on using this paradigm, an ideal method for evaluating and expressing measurement uncertainty should efficiently provide such an interval (or region), possibly under additional constraints, with a coverage probability or level of confidence that corresponds to the required level of certainty, typically with a specified coverage level, such as $95\%$.  Generally, increasing the coverage level results in a wider coverage interval or a larger coverage region.}



{
One of the alternative methods to the Monte Carlo method (MCM) proposed in~\cite{GUMS1, GUMS2} for deriving the probability distribution of the measured quantity and assessing measurement uncertainty is the Characteristic Function Approach
\cite{witkovsky2017}.} This approach uses characteristic functions to fully determine the behavior and properties of probability distributions of random variables. Although characteristic functions have a fundamental role in several fields of mathematics, probability theory, statistics and engineering, they can be challenging to work with, particularly when it comes to inverting them to obtain the cumulative distribution function (CDF) or the probability distribution function (PDF) of the corresponding random variable and subsequently deriving the corresponding coverage interval with a specified level of confidence. To address these challenges, various tools and methods have been developed for evaluating, combining and inverting characteristic functions.

{The CFA is primarily used to assess the probability distribution of a quantity that depends linearly on multiple input quantities with different probability distributions.} However, it can also be applied to nonlinear measurement models if the characteristic function of the output quantity can be derived or by approximating the nonlinear function linearly to derive the characteristic function of the output quantity. This linear approximation is based on the first-order Taylor series expansion of the nonlinear function around the expected value or the best estimate of the input quantities. The CFA is capable of handling any type of probability distribution, including non-Gaussian and asymmetric ones, as long as the characteristic functions of the input quantities are known. The CFA obtains the characteristic function of the output quantity by applying a transformation rule to the characteristic functions of the input quantities.

In fact, the characteristic function of a weighted sum of independent random variables is simple to derive if the measurement model is linear, i.e.,
\begin{equation}\label{eq02}
Y = f(X_1,\dots,X_n) = c_1 X_1 + \cdots + c_n X_n,
\end{equation}
for some known constants $c_1, \dots , c_n$ (the sensitivity coefficients) and the input quantities $X_i$ are mutually independent random variables with known distributions. In such situations, the CF of the output quantity $Y$ is given as
\begin{equation}\label{eq03}
\operatorname{cf}_{Y}(t) = \operatorname{cf}_{X_1 }(c_1t) \times \cdots \times \operatorname{cf}_{X_n }(c_nt),
\end{equation}
where by $\operatorname{cf}_{X_i }(t)$ we denote the (known) CFs of the input quantities $X_i$. 

The probability distribution function and the cumulative distribution function of the output quantity $Y$ can be obtained by numerically inverting its characteristic function, which can be efficiently calculated, for example, using a simple trapezoidal quadrature or other advanced quadrature rules as is the adaptive Gauss--Kronod quadrature, depending on the complexity and oscilatory properties of the integrand function; see \cite{witkovsky2016,witkovsky2023}. From the cumulative distribution function, the quantile function and coverage intervals with a specified coverage level can be derived by selecting appropriate quantiles of the~distribution.


{The CFA compares favorably with the MCM, which estimates the probability distribution of the output quantity by generating random samples from the input quantities and applying the measurement model and the law of propagation of distributions \cite{GUMS1,GUMS2}. It offers several advantages, including accuracy ({The proper error analysis of the approaches used for numerical inversion has been studied in the literature for several particular cases. In general, it is still an open problem which depends on the specific inversion algorithm and the properties of the characteristic function. Shepard in \cite{shephard1991b} analyzed the error caused by approximating the required inversion integral under specific assumptions. The author suggested that using the trapezium rule to approximate the integral makes it challenging to manage the induced integration error. Therefore, a simpler Riemann sum has been employed instead, resulting in a much simpler numerical integration error. This formulation has been successfully applied to solve the problem of finding the exact distribution function of a quadratic form in normal variables with adequate control of the numerical error, as demonstrated by Davies in \cite{davies1980}.}), flexibility, efficiency and repeatability, making it a valuable method for assessing measurement uncertainty. However, it faces challenges in finding an explicit expression for the characteristic function of the output quantity and inverting it to obtain the CDF, PDF or other statistics. It is worth noting that deriving the joint characteristic function of a multivariate distribution with stochastically dependent input quantities and numerically inverting such a characteristic function can be~challenging. 
}

\section{Characteristic Function of $q$-Gaussian Distribution}
\label{sec3}
Here we present the derivation of the characteristic function for the $q$-Gaussian distribution as well as the characteristic function for a linear combination of independent $q$-Gaussian random variables, $Y = \sum_{k=1}^n c_k X_k$, where $X_k$ are independent $q$-Gaussian random variables possibly with different parameters and $c_k$ are known constant coefficients.

The probability density function of the Tsallis $q$-Gaussian distribution is given by:
\begin{equation}
f(x) = C_{q,\sigma}  \left[1 - (1-q) \frac{1}{2} \left(\frac{x-\mu}{\sigma}\right)^2 \right]_+^{\frac{1}{1-q}},
\end{equation}
where $C_{q,\sigma} $ is a normalization constant, $\mu$ (real) and $\sigma > 0$ are the location and scale parameters of the distribution and $[z]_+ = \max(0,z)$. The shape parameter $q<3$ controls the degree of non-extensivity of the system and the distribution reduces to the Gaussian distribution when $q = 1$.
 
The parametrization used in this paper is consistent with 
\href{https://reference.wolfram.com/language/ref/TsallisQGaussianDistribution.html}{Wolfram Mathematica}, where $\sigma > 0$ represents the scale parameter. It is important to note that this differs from the parametrization defined by Tsallis---for details see also
\href{https://en.wikipedia.org/wiki/Q-Gaussian_distribution}{Wikipedia}---where 
\begin{equation}
f(x) = C_{q,\beta}  \left[1 - (1-q) \beta \left(x-\mu\right)^2 \right]_+^{\frac{1}{1-q}}, 
\end{equation} 
which uses $\beta > 0$ as the rate parameter, such that $\beta = 1/(2\sigma^2)$ or $\sigma = \sqrt{1/(2\beta)}$. 

We use $X \sim \mathit{TQG}(\mu, \sigma, q)$ to denote a random variable with a Tsallis $q$-Gaussian distribution, where $\mu \in \mathbb{R}$ is the location parameter, $\sigma > 0$ is the scale parameter and $q < 3$ is the shape parameter known as the Tsallis $q$-index. The distribution $\mathit{TQG}(\mu, \sigma, q)$ is parametrized such that it is a normal distribution with mean $\mu$ and variance $\sigma^2$ when $q=1$, i.e.,~$X \sim \mathit{N}\left(\mu, \sigma^2\right)$, or

\begin{equation}\label{eqnXN}
X = \mu + \sigma Z, 
\end{equation} 
where $Z\sim\mathit{N}\left(0, 1\right)$ is a random variable with the standard normal distribution.

Moreover, for $q<1$, $X$ is a bounded random variable with its distribution proportional to a scaled and shifted symmetric beta distribution. In particular, we obtain
\begin{equation}\label{eqnXB}
X = \mu + \sigma \sqrt{\frac{2}{1-q}}\left(2B-1\right), 
\end{equation} 
where $B\sim\mathit{Beta}\left(\theta,\theta\right)$ is a beta distributed random variable with both shape parameters equal to $\theta = \frac{2-q}{1-q}$. The support of this distribution is limited to 

\begin{equation}
\left\langle \mu - \sigma \sqrt{\frac{2}{1-q}}, \mu + \sigma \sqrt{\frac{2}{1-q}} \right\rangle. 
\end{equation}

For $1<q<3$, the distribution is formally proportional to a scaled and shifted Student $t$-distribution, which is often preferred over the Gaussian distribution due to its heavy tails, meaning that it has a higher probability of extreme events. In particular, we obtain
\begin{equation}\label{eqnXT}
X = \mu + \sigma \sqrt{\frac{2}{3-q}} T, 
\end{equation} 
where $T\sim\mathit{t}\left(\nu\right)$ is a Student $t$-distributed random variable with $\nu = \frac{3-q}{q-1}$ degrees of freedom.

It is worth noting that as $q$ approaches 1 from above, $\nu$ approaches infinity (i.e.,~normal distribution). For $q \geq \frac{5}{3}$, the second moments are infinite or do not exist. For $2 \leq q<3$, the $q$-Gaussian distribution is located in a region of extremely heavy tails where the moments of the distribution may not exist or may not be well-defined. If $q=2$, the $q$-Gaussian distribution is proportional to a scaled and shifted Cauchy distribution; otherwise, it is a $t$-distribution with fractional degrees of freedom, $0 < \nu < 1$. Specifically, as $q$ approaches 3, $\nu\to 0$ and as $q$ approaches 2, $\nu\to 1$.

The main result of the paper is presented in Theorem~1 and was obtained through symbolic computation using Wolfram Mathematica and validated through comparison with the characteristic functions of the Student $t$ and symmetric beta distributions;  see~\cite{witkovsky2001,witkovsky2023}.

\begin{theorem}[Characteristic function of the Tsallis $q$-Gaussian]
The characteristic function of the Tsallis $q$-Gaussian distribution with the parameters $\mu\in \mathbb{R}$, $\sigma >0$ and $q < 3$ is defined as
\begin{equation}
		\operatorname{cf}_{\mathit{TQG}(\mu,\sigma,q)}(t) = \exp(\mathrm{i} t \mu) \times  \operatorname{cf}_{\mathit{TQG}(0,1,q)}(\sigma t), 
\end{equation}
where $\mathrm{i} = \sqrt{-1}$ and $\operatorname{cf}_{\mathit{TQG}(0,1,q)}(t)$ is CF of the standard Tsallis $q$-Gaussian distribution,
\begin{itemize}
	\item 
for $q<1$ defined as
\begin{align}		
\operatorname{cf}_{\mathit{TQG}(0,1,q)}(t) &= {\,_0\operatorname{F}_1}\left(\theta + \frac{1}{2}, -\frac{1}{4} (a t)^{2}\right) 
   =  2^{\theta - \frac{1}{2}} \Gamma\left(\theta + \frac{1}{2}\right)  (a t)^{-(\theta + \frac{1}{2})}  \operatorname{J}_{\theta - \frac{1}{2}}\left(a t\right)\cr 
	&=   \operatorname{cf}_{\mathit{BetaSymmetric}(\theta)}(a t),
\end{align}
 where $a = \sqrt{\frac{2}{1-q}}$, $\theta = \frac{2-q}{1-q}$, ${\,_0\operatorname{F}_1}\left(b,z\right)$ is a confluent hypergeometric function, $\Gamma\left(z\right)$ is a gamma function, $\operatorname{J}_{\nu}\left(z\right)$ is a Bessel function of the first kind and $\operatorname{cf}_{\mathit{BetaSymmetric}(\theta)}(t)$ denotes CF of the symmetric beta distribution with the parameter $\theta$ and the support on the \linebreak interval $\langle -1,1\rangle$, 
\item 
for $q=1$ defined as
\begin{align}
\operatorname{cf}_{\mathit{TQG}(0,1,q)}(t) &=  \exp\left(-\frac{t^2}{2}\right) =  \operatorname{cf}_{\mathit{Normal}\left(0,1\right)}(t). 
\end{align}
\item 
for $1< q<3$ defined as
\begin{align}
\operatorname{cf}_{\mathit{TQG}(0,1,q)}(t) &= \frac{\left( b \sqrt{\nu} |t| \right)^{\frac{\nu}{2}} \operatorname{K}_{\frac{\nu}{2}}\left(b \sqrt{\nu} |t|\right) } { 2^{\frac{\nu}{2} - 1} \Gamma\left(\frac{\nu}{2}  \right)} =  \operatorname{cf}_{\mathit{Student}(\nu)}(bt),
\end{align}
where $b = \sqrt{\frac{2}{3-q}}$, $\nu = \frac{3-q}{q-1}$, $\operatorname{K}_{\nu}\left(z\right)$ is a Bessel function of the second kind and $\operatorname{cf}_{\mathit{Student}(\nu)}(t)$ denotes CF of the Student $t$-distribution with $\nu$ degrees of freedom with $\nu > 0$. 
\end{itemize}
\end{theorem}

\begin{corollary}[Characteristic function of a linear combination of the Tsallis $q$-Gaussians]
Let $\operatorname{cf}_{Y}(t)$ denote the characteristic function of a linear combination of independent random variables $Y = \sum_{k=1}^n c_k X_k$, where $c_k$ are real coefficients and $X_k$ are independent Tsallis $q$-Gaussians with characteristic functions $\operatorname{cf}_{X_k} = \operatorname{cf}_{\mathit{TQG}(\mu_k,\sigma_k,q_k)}(t)$ for $k=1,\ldots,n$. Then, the characteristic function of $Y$ can be expressed as follows:

\begin{equation}
\operatorname{cf}_{Y}(t) = \prod_{k=1}^n \operatorname{cf}_{X_k}\left(c_kt\right) = \prod_{k=1}^n \operatorname{cf}_{\mathit{TQG}(\mu_k,\sigma_k,q_k)}\left(c_kt\right).
\end{equation}

\end{corollary}
This formula allows us to compute the characteristic function of $Y$ by taking the product of the characteristic functions of each $X_k$, evaluated at $c_kt$.

\section{CharFunTool: The Characteristic Functions Toolbox}
\label{sec4}

While the literature presents various techniques and tools for evaluating, combining and inverting characteristic functions, few reliable and efficient software implementations exist. One notable implementation, still under continuous development, is the MATLAB toolbox
\href{https://github.com/witkovsky/CharFunTool}{\textbf{CharFunTool}} \cite{witkovsky2023}, which offers a range of characteristic functions for various probability distributions and algorithms for their manipulation and inversion. The toolbox can evaluate, combine and invert characteristic functions of different types of probability distributions, including continuous, discrete, circular and bivariate distributions, as well as mixture and empirical distributions. With over 60 characteristic functions available, the toolbox also provides a wide variety of other symmetric and non-negative probability distributions, such as Beta, Chi-Square, Exponential, Gamma, Normal, Poisson, Rayleigh, Student, Weibull and many others. 

The algorithm
\href{https://github.com/witkovsky/CharFunTool/blob/master/CF_Repository/cf_TsallisQGaussian.m}{\textbf{cf\_TsallisQGaussian}} was implemented into the repository of the characteristic functions of the toolbox. The algorithm evaluates the characteristic function of the random variable $Y = \sum_{k=1}^n c_k X_k$, where $c_k$ are real coefficients and $X_k$ are independent Tsallis $q$-Gaussians with arbitrary parameters for $k=1,\ldots,n$. In general, the toolbox offers a unique and straightforward approach to evaluating, combining and inverting combined characteristic functions. 

The well-known method for inverting characteristic functions that satisfies conditions specified  in \cite{wendel1961,shephard1991a} is the Gil-Pelaez formula, which was proposed by Gil-Pelaez in 1951 \cite{gilpelaez1951}. This formula allows one to invert the characteristic function by using numerical integration techniques. The formula states that for a univariate random variable $X$, if $x$ is a continuity point of its cumulative distribution function, then
\begin{equation}\label{eq04}
\operatorname{cdf}_X(x) = \frac{1}{2} - \frac{1}{\pi} \int_0^\infty \frac{\operatorname{Im}\left[e^{-\mathrm{i}tx}\operatorname{cf}_X(t)\right]}{t}\, dt.
\end{equation}

Moreover,
\begin{equation}\label{eq05}
\operatorname{pdf}_X(x) = \frac{1}{\pi} \int_0^\infty {\operatorname{Re}\left[e^{-\mathrm{i}tx}\operatorname{cf}_X(t)\right]}\, dt,
\end{equation}
where $\operatorname{cdf}_X$ is the CDF of $X$, $\operatorname{pdf}_X$ is the PDF of $X$, $\operatorname{cf}_X$ is the characteristic function of $X$ and $\operatorname{Re}$ and $\operatorname{Im}$ denote the real and imaginary parts of complex numbers, respectively. However, the use of the formulae may be affected by numerical instability and slow convergence in cases where the characteristic function or the integrands exhibits oscillatory or singular~behavior.

There are many alternative methods and improvements available for computing the numerical inversion of the characteristic function. Several of these methods have been directly implemented or have inspired the development of inversion algorithms in the \href{https://github.com/witkovsky/CharFunTool}{\textbf{CharFunTool}}, namely \cite{abate2004,bakhvalov1968,chourdakis2005,cohen2000,
davies1973,evans1999,hurlimann2013,imhof1961,mijanovic2023,shephard1991a,shephard1991b,sidi2012,talbot1979,waller1995,witkovsky2016}. 

The toolbox offers a range of algorithms for numerically inverting characteristic functions of univariate and bivariate probability distributions, which can be categorized into two categories: univariate inversion algorithms and bivariate inversion algorithms.
	
\subsection*{Inversion Algorithms}

Univariate inversion algorithms are used to obtain the probability distribution function, cumulative distribution function, quantile function (QF) or probability mass function (PMF) of a univariate random variable, given its characteristic function. Various algorithms for univariate inversion include:

\begin{itemize}
\item 
\href{https://github.com/witkovsky/CharFunTool/blob/master/CF_InvAlgorithms/cf2DistGP.m}{\textbf{cf2DistGP}}: This algorithm uses the Gil-Pelaez inversion formulae (\ref{eq04}) and (\ref{eq05}) to return the PDF, CDF or QF of a univariate distribution. It also provides an option to use different integration algorithms such as Riemann sum, trapezoidal rule or adaptive Gauss--Kronrod quadrature rule for more efficient calculations.
\item 
\href{https://github.com/witkovsky/CharFunTool/blob/master/CF_InvAlgorithms/cf2DistGPR.m}{\textbf{cf2DistGPR}}: This algorithm returns the PDF, CDF or QF of a univariate distribution using the Gil-Pelaez inversion formulae with the Riemann sum integration method. For more details, see, e.g., \cite{shephard1991a}.
\item 
\href{https://github.com/witkovsky/CharFunTool/blob/master/CF_InvAlgorithms/cf2DistGPT.m}{\textbf{cf2DistGPT}}: This algorithm returns the PDF, CDF or QF of a univariate distribution using the Gil-Pelaez inversion formulae with the trapezoidal rule integration method. For more details, see, e.g., \cite{davies1973,shephard1991a,waller1995,witkovsky2016}.
\item 
\href{https://github.com/witkovsky/CharFunTool/blob/master/CF_InvAlgorithms/cf2DistGPA.m}{\textbf{cf2DistGPA}}: This algorithm returns the PMF of a discrete univariate distribution using the Gil-Pelaez inversion formulae with the adaptive Gauss--Kronrod quadrature rule integration method. For more details, see, e.g., \cite{cohen2000,sidi2012}.
\item 
\href{https://github.com/witkovsky/CharFunTool/blob/master/CF_InvAlgorithms/cf2DistBTAV.m}{\textbf{cf2DistBTAV}}: This algorithm uses the Bromwich--Talbot--Abate--Valko (BTAV) method to return the CDF, PDF or QF of a non-negative univariate distribution specified by a given characteristic function. For more details, see, e.g., \cite{abate2004,evans1999}.
\item 
\href{https://github.com/witkovsky/CharFunTool/blob/master/CF_InvAlgorithms/cf2DistBV.m}{\textbf{cf2DistBV}}: This algorithm returns the CDF, PDF or QF of a univariate distribution using the Gil-Pelaez inversion formulae and the Bakhvalov--Vasileva method \cite{bakhvalov1968}.
\item 
\href{https://github.com/witkovsky/CharFunTool/blob/master/CF_InvAlgorithms/cf2DistFFT.m}{\textbf{cf2DistFFT}}: This algorithm returns the CDF, PDF or QF of a univariate distribution using the Fast Fourier Transform (FFT) algorithm. For more details, see, e.g., \cite{chourdakis2005,hurlimann2013}.
\end{itemize}

The bivariate inversion algorithm takes a characteristic function of a bivariate random vector and returns the joint PDF or CDF of the corresponding random vector:
\begin{itemize}
\item 
\href{https://github.com/witkovsky/CharFunTool/blob/master/CF_InvAlgorithms/cf2Dist2D.m}{\textbf{cf2Dist2D}}: This returns the CDF, PDF, quantile function or random numbers of a bivariate distribution using the Gil-Pelaez inversion formulae with Fourier integrals calculated using the simple Riemann sum quadrature method suggested by Shephard in 1991~\cite{shephard1991b}. See also \cite{mijanovic2023}.
\end{itemize}

Note that the choice of algorithm may depend on the properties of the input characteristic function, such as its smoothness, oscillatory behavior and integrability.



\section{Numerical Examples}
\label{sec5}

{
The examples in this section do not include specific measurement uncertainty problems. Instead, they illustrate the application and implementation of the CFA as an alternative tool for mathematical computation of the probability distribution of the output quantity in linear measurement models with $q$-Gaussian input variables, with potential applications for measurement uncertainty analysis.
}

For illustration, here we present the application of the \href{https://github.com/witkovsky/CharFunTool}{\textbf{CharFunTool}} algorithms for evaluating PDF, CDF and specified quantiles of the distribution of an output quantity. For each of those four examples, we consider a linear combination of independent $q$-Gaussian random variables with different parameters specified as $Y = c_1 X_1 + c\dots + c_n X_n$, where the coefficients $c_i$ and the characteristic functions $\operatorname{cf}_{X_i}$ are given and this demonstrates how the CFA can be used to obtain the PDF, CDF and QF of the output quantity. The examples cover a range of scenarios, including situations where all $q$-index parameters are less than 1, where the $q$-index parameters are a mixture of positive and negative values, where the Tsallis parametrization is used and where extreme values of $q$ are considered.

\subsection{Example 1: PDF/CDF/QF of a Linear Combination of $q$-Gaussian RVs with Small $q$-Indices, $q\leq 0$}

\begin{figure}
\includegraphics[width=0.5\textwidth]{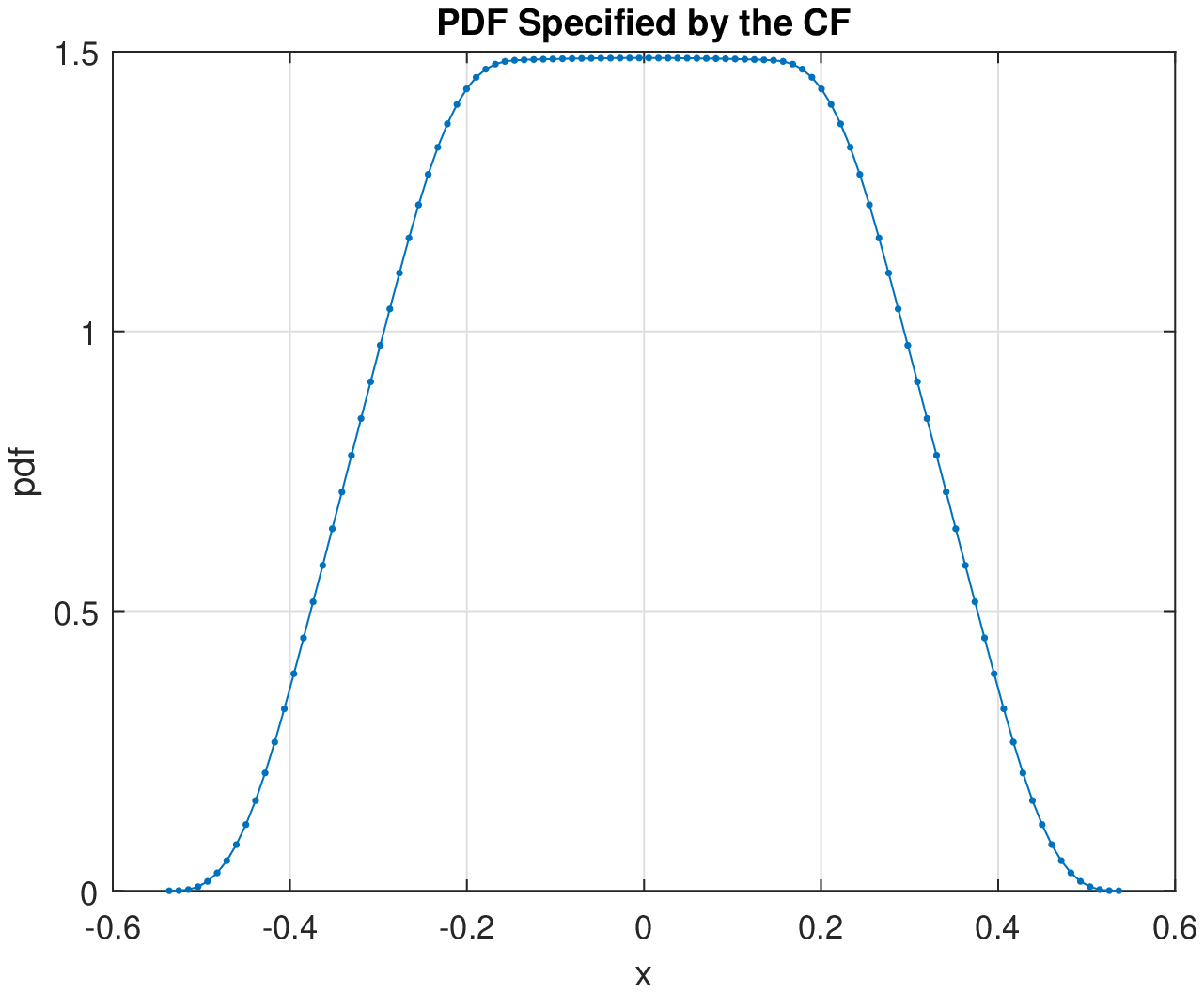}
\includegraphics[width=0.5\textwidth]{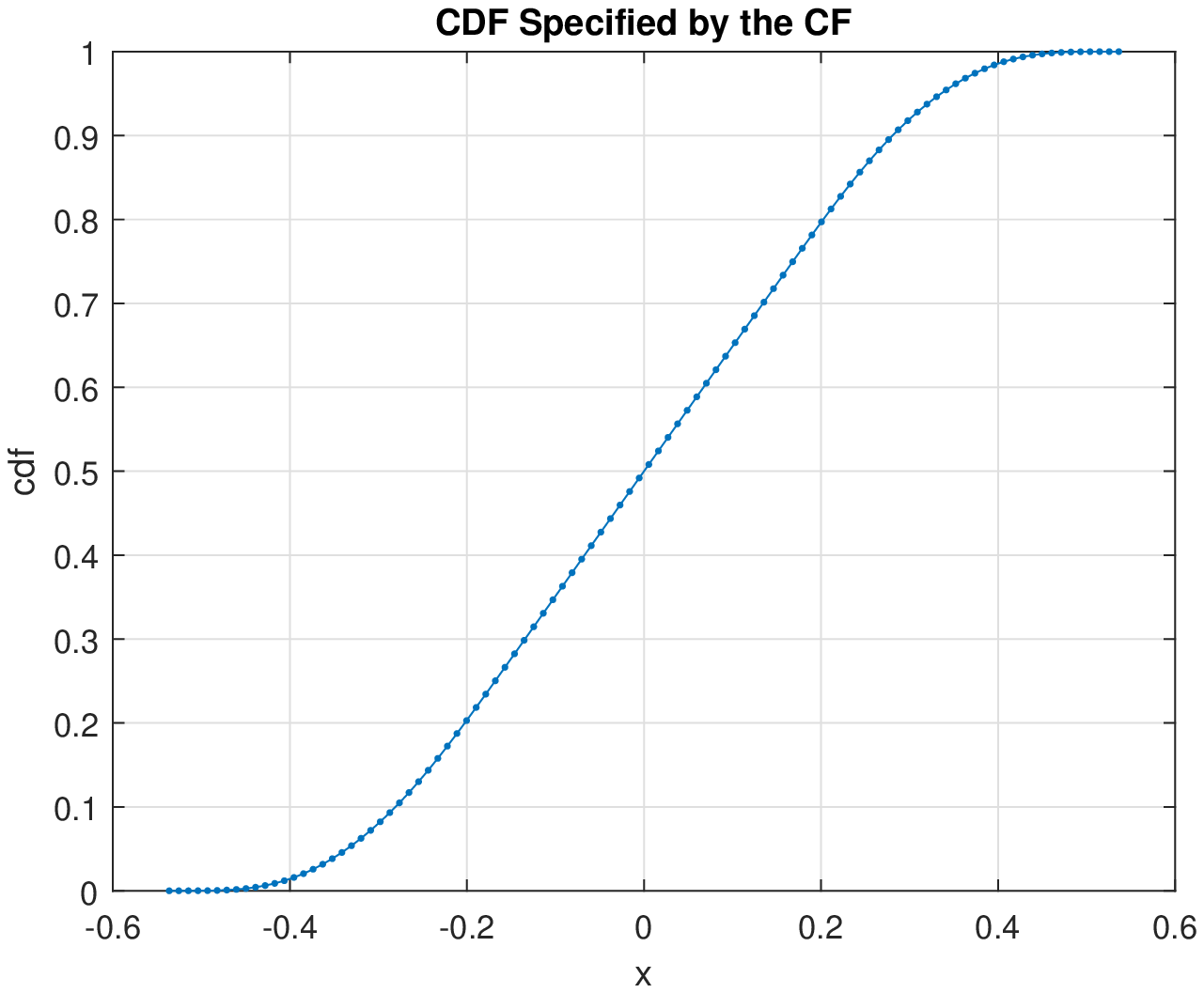}
\begin{lstlisting}
%% EXAMPLE 1: PDF/CDF/QF of a linear combination of independent q-Gaussians 
%  RVs with all q < 1

%  Characteristic Function Approach (CFA): 
mu     = [0 0 0];
sigma  = [3 2 1];
q      = [-100 -10 0];
coef   = [0.8 0.15 0.05];
cf     = @(t) cf_TsallisQGaussian(t,mu,sigma,q,coef);
clear options
options.N    = 2^10;
options.xMin = sum(mu - sigma.*sqrt(2./(1-q)) .* coef);
options.xMax = sum(mu + sigma.*sqrt(2./(1-q)) .* coef);
x      = linspace(options.xMin,options.xMax)';
prob   = [0.025 0.975];
result = cf2DistGP(cf,x,prob,options);
disp(result)
\end{lstlisting}
	\caption{Characteristic 
 function approach  using the MATLAB algorithms from \href{https://github.com/witkovsky/CharFunTool}{\textbf{CharFunTool}}. PDF, CDF and QF computed from the characteristic function of a linear combination of $q$-Gaussian RVs, $Y = 0.8 X_1 + 0.15 X_2 + 0.05 X_3$, where $X_1 \sim \mathit{TQG}(0,3,-100)$, $X_2 \sim \mathit{TQG}(0,2,-10)$ and $X_3 \sim \mathit{TQG}(0,1,0)$ are independent random variables. The $95\%$ coverage interval derived by CFA is $\operatorname{CI}_{CFA} = [-0.3751, 0.3751]$.
	\label{fig00}}
\end{figure}

Here we consider a linear combination of three independent $q$-Gaussian random variables, 
\begin{equation}
Y = 0.8 X_1 + 0.15 X_2 + 0.05 X_3, 
\end{equation}
where $X_1 \sim \mathit{TQG}(0,3,-100)$, $X_2 \sim \mathit{TQG}(0,2,-10)$ and $X_3 \sim \mathit{TQG}(0,1,0)$ are independent random variables. We used the \href{https://github.com/witkovsky/CharFunTool}{\textbf{CharFunTool}} and the characteristic function of $Y$ to compute its PDF, CDF and QF. Figure~\ref{fig00} shows the graphs of the PDF and CDF along with the MATLAB code used to evaluate the result. The CFA was used to derive a $95\%$ coverage interval, which is given by $\operatorname{CI}_{CFA} = [-0.3751, 0.3751]$.

Alternatively, we can use MCM with (\ref{eqnXB}) to derive the result by generating $N=$ 100,000 realizations of the $X_i$  to obtain $N$ realizations of $Y$. By sorting them, we can then derive the required coverage interval as $\operatorname{CI}_{MCM} = [Y_{sort}( \lfloor N \times 0.025 \rfloor ), Y_{sort}( \lceil N \times (1-0.025)\rceil )]$. Using this method, we obtain $\operatorname{CI}_{MCM} = [-0.3762, 0.3753]$ for comparison. Here, we can see that the coverage intervals derived using CFA and MCM are very close, with only a small difference between them.

\subsection{Example 2: PDF/CDF/QF of a Linear Combination of $q$-Gaussian RVs with Different Types of $q$-Indices, $q \leq 1.5$}

\begin{figure}
\includegraphics[width=0.5\textwidth]{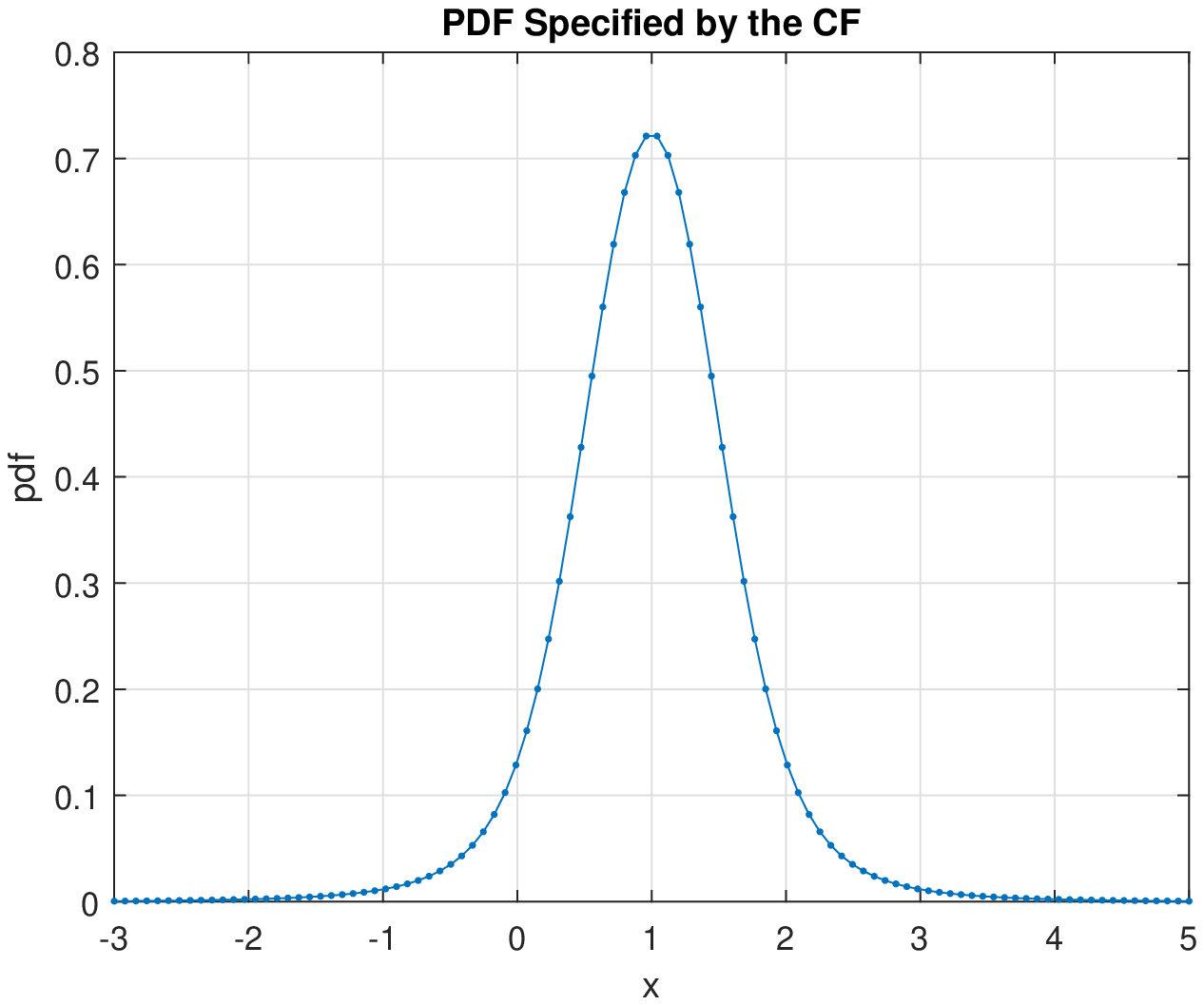}
\includegraphics[width=0.5\textwidth]{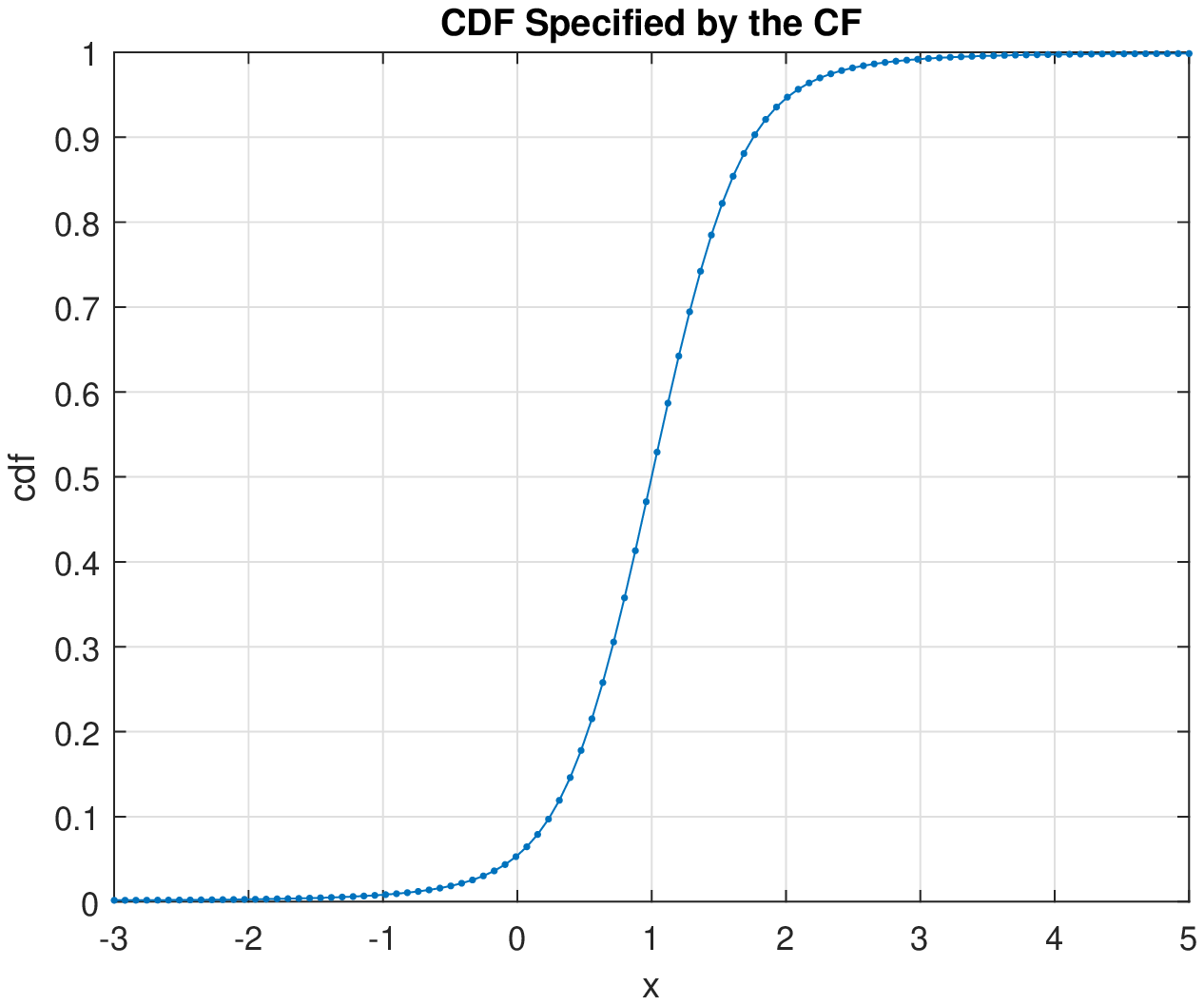}
\begin{lstlisting}
%% EXAMPLE 2: PDF/CDF/QF of a linear combination of independent q-Gaussians 
%  RVs with different types of q-indices: q < 1 and q > 1

%  Characteristic Function Approach (CFA):
mu     = [0 1 2];
sigma  = [1 1 1];
q      = [-1 0.5 1.5];
coef   = [1 1 1]/3;
cf     = @(t) cf_TsallisQGaussian(t,mu,sigma,q,coef);
clear options
options.N = 2^10;
x      = linspace(-3,5)';
prob   = [0.025 0.975];
result = cf2DistGP(cf,x,prob,options);
disp(result)
\end{lstlisting}
	\caption{Characteristic 
 function approach  using the MATLAB algorithms from \href{https://github.com/witkovsky/CharFunTool}{\textbf{CharFunTool}}. PDF, CDF and QF computed from the characteristic function of a linear combination of $q$-Gaussian RVs, $Y = \frac{1}{3} X_1 + \frac{1}{3}  X_2 + \frac{1}{3}  X_3$, where $X_1 \sim \mathit{TQG}(0,1,-1)$, $X_2 \sim \mathit{TQG}(1,1,0.5)$ and $X_3 \sim \mathit{TQG}(2,1,1.5)$ are independent random variables. The $95\%$ coverage interval derived via CFA is $\operatorname{CI}_{CFA} = [-0.3409, 2.3409]$.
		\label{fig01}}
\end{figure}

Consider a linear combination 
\begin{equation}
Y = \frac{1}{3} X_1 + \frac{1}{3} X_2 + \frac{1}{3} X_3,
\end{equation}
where $X_1 \sim \mathit{TQG}(0,1,-1)$, $X_2 \sim \mathit{TQG}(1,1,0.5)$ and $X_3 \sim \mathit{TQG}(2,1,1.5)$ are independent random variables with different types of $q$-indices. Figure~\ref{fig01} shows the PDF and CDF graphs along with the MATLAB code used to evaluate the result. The CFA was used to derive a $95\%$ coverage interval, which is given by $\operatorname{CI}_{CFA} = [-0.3409, 2.3409]$.

We used MCM with (\ref{eqnXB}) and (\ref{eqnXT}) to derive the result by generating $N$ = 100,000 realizations of  $X_i$ to obtain $N$ realizations of $Y$. By sorting the realizations of $Y$, we obtained $\operatorname{CI}_{MCM} = [-0.3248, 2.3375]$ for comparison. As before, we can see that the coverage intervals derived using CFA and MCM are very close, with only a small difference between~them.

\subsection{Example 3: PDF/CDF/QF of a Linear Combination of $q$-Gaussian RVs with Different Types of $q$-Indices, $q \leq 2$ and  Using the Tsallis Parametrization}

\begin{figure}
\includegraphics[width=0.5\textwidth]{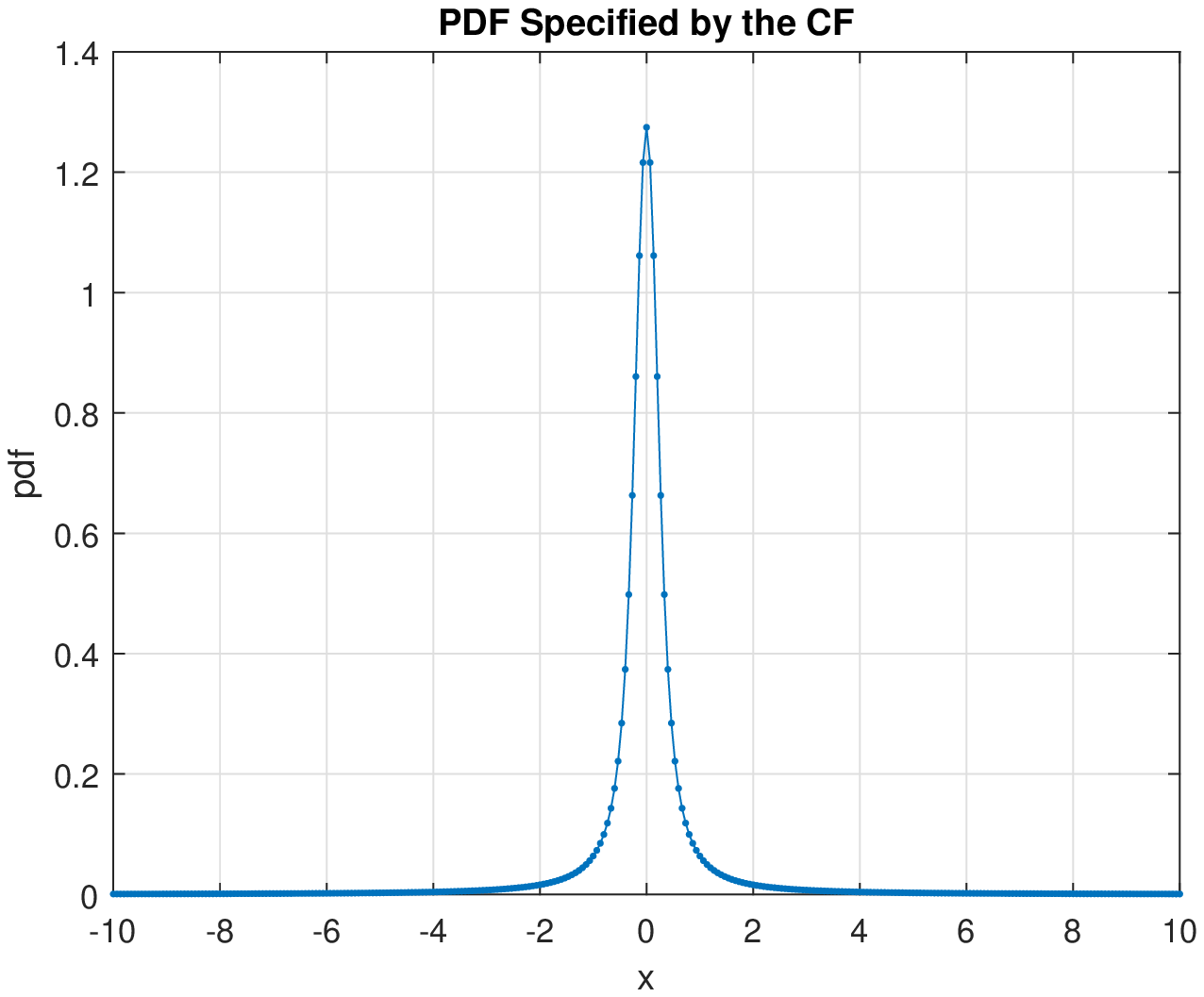}
\includegraphics[width=0.5\textwidth]{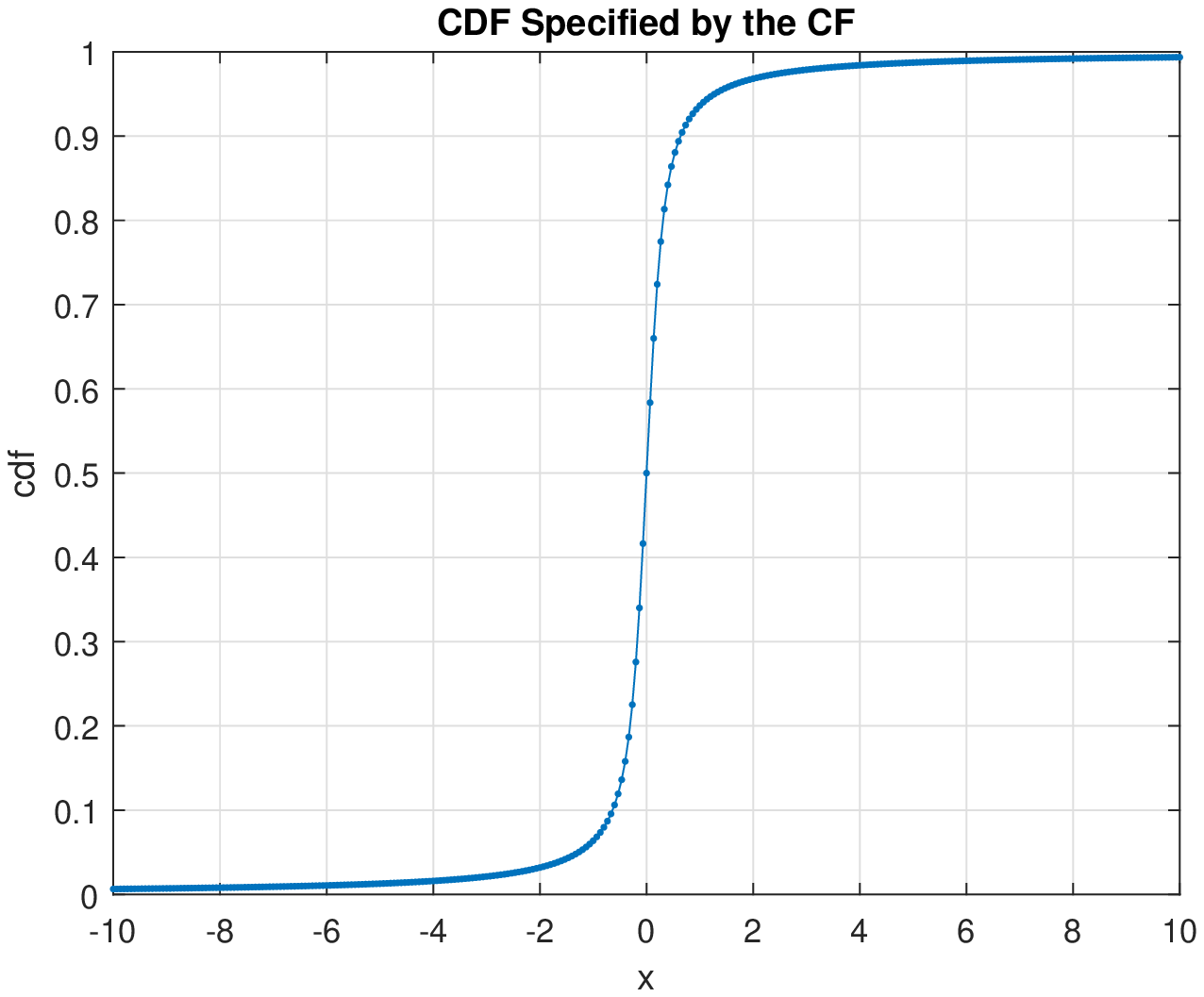}
\begin{lstlisting}
%% EXAMPLE 3: PDF/CDF/QF of a linear combination of independent q-Gaussians 
%  RVs with different types of q-indices: q < 1 and q > 1
%  with using the Tsallis parametrization

%  Characteristic Function Approach (CFA):
mu     = [0 0 0 0 0];
beta   = [5 4 3 2 1];
sigma  = sqrt(1./(2*beta));
q      = [-5 -1 0 1 2];
coef   = [1 1 1 1 1]/5;
cf     = @(t) cf_TsallisQGaussian(t,mu,sigma,q,coef);
clear options
options.N = 2^14;
x      = linspace(-10,10,301);
prob   = [0.025 0.975];
result = cf2DistGP(cf,x,prob,options);
disp(result)
\end{lstlisting}
	\caption{Characteristic function approach  using the MATLAB algorithms from \href{https://github.com/witkovsky/CharFunTool}{\textbf{CharFunTool}}. PDF, CDF and QF computed from the characteristic function of a linear combination of $q$-Gaussian RVs, $Y = \frac{1}{5} X_1 + \frac{1}{5}  X_2 + \frac{1}{5}  X_3 + \frac{1}{5}  X_4+ \frac{1}{5}  X_5$, where $X_1 \sim \mathit{TQG}(0,\sqrt{{1}/{10}},-5)$, $X_2 \sim \mathit{TQG}(0,\sqrt{{1}/{8}},-1)$, $X_3 \sim \mathit{TQG}(0,\sqrt{{1}/{6}},0)$, $X_4 \sim \mathit{TQG}(0,\sqrt{{1}/{4}},1)$ and\linebreak $X_5 \sim \mathit{TQG}(0,\sqrt{{1}/{2}},2)$ are independent random variables specified from the given Tsallis parametrization. The $95\%$ coverage interval derived by CFA is $\operatorname{CI}_{CFA} = [ -2.5469 , 2.5469]$.	\label{fig02}}
\end{figure}

Consider a linear combination 
\begin{equation}
Y = \frac{1}{5} X_1 + \frac{1}{5}  X_2 + \frac{1}{5}  X_3 + \frac{1}{5}  X_4+ \frac{1}{5}  X_5,
\end{equation}
where $X_1 \sim \mathit{TQG}(0,\sqrt{{1}/{10}},-5)$, $X_2 \sim \mathit{TQG}(0,\sqrt{{1}/{8}},-1)$, $X_3 \sim \mathit{TQG}(0,\sqrt{{1}/{6}},0)$, \linebreak$X_4 \sim \mathit{TQG}(0,\sqrt{{1}/{4}},1)$ and $X_5 \sim \mathit{TQG}(0,\sqrt{{1}/{2}},2)$ are independent random variables defined  using the Tsallis parametrization. Figure~\ref{fig02} shows the PDF and CDF graphs along with the MATLAB code used to evaluate the result. The CFA was used to derive a $95\%$ coverage interval, which is given by $\operatorname{CI}_{CFA} = [-2.5469 , 2.5469]$.

We used MCM with (\ref{eqnXB}), (\ref{eqnXN}) and (\ref{eqnXT}) to derive the result by generating $N$ = 100,000 realizations of the $X_i$ to obtain $N$ realizations of $Y$. By sorting the realizations of $Y$, we obtained $\operatorname{CI}_{MCM} = [ -2.5482 , 2.5018]$ for comparison. As before, we can see that the coverage intervals derived using CFA and MCM are very close, with only a small difference between~them.

\subsection{Example 4: CDF of a Linear Combination of $q$-Gaussian RVs  Including Large $q$-Indices with $\max{q} = 2.9$}

We now consider an extreme case, where we have a linear combination of independent random variables $X_1$, $X_2$ and $X_3$ given by:
\begin{equation}
Y = \frac{1}{3} X_1 + \frac{1}{3} X_2 + \frac{1}{3} X_3,
\end{equation}
where $X_1 \sim \mathit{TQG}(0,1,0)$, $X_2 \sim \mathit{TQG}(0,0.5,1)$ and $X_3 \sim \mathit{TQG}(0,1,2.9)$ are independent random variables.

{
We used MATLAB code to evaluate the specified CDF values and the $95\%$ coverage interval for $Y$, as shown in Figure~\ref{fig03}. The exact $95\%$ coverage interval derived by CFA is $\operatorname{CI}_{CFA} = [-9.1540 \times 10^{22} , 9.1540\times 10^{22}]$.

\begin{figure}
\begin{lstlisting}
%% EXAMPLE 4: CDF and 95% coverage interval 
%  of a linear combination of independent q-Gaussians 
%  RVs with different types of q-indices 
%  including extreme values of q (here max q = 2.9) 
%  computed with using cf2CDF_GPA and cf2QF_GPA

mu     = [0 0 0];
sigma  = [1 0.5 0.1];
q      = [ 0 1 2.9];
coef   = [1 1 1]/3;
cf     = @(t) cf_TsallisQGaussian(t,mu,sigma,q,coef);
clear options
options.isAccelerated = true; 

% 95% Coverage interval CI 
CI_Low = cf2QF_GPA(cf,0.025,options);
CI_Upp = cf2QF_GPA(cf,0.975,options);
CI     = [CI_Low, CI_Upp];
disp (CI)

% CDF values at specified arguments of y
y      = [1e+10 1e+20 1e+30 1e+40 1e+50 1e+60 1e+70 1e+80 1e+90]';
cdf    = cf2CDF_GPA(cf,y,options);
Table  = table(y,cdf)

     y         cdf  
  --------  ---------
    1e+10    0.87974
    1e+20    0.96421
    1e+30    0.98935
    1e+40    0.99683
    1e+50    0.99906
    1e+60    0.99972
    1e+70    0.99992
    1e+80    0.99998
    1e+90    0.99999
		
\end{lstlisting}
\caption{Characteristic Function Approach based on using the MATLAB algorithms from \href{https://github.com/witkovsky/CharFunTool}{\textbf{CharFunTool}}. CDF values and the $95\%$ coverage interval  computed from the characteristic function of a linear combination of $q$-Gaussian RVs, \mbox{$Y = \frac{1}{3} X_1 + \frac{1}{3}  X_2 + \frac{1}{3}  X_3$}, where $X_1 \sim \mathit{TQG}(0,1,0)$, $X_2 \sim \mathit{TQG}(0,0.5,1)$ and $X_3 \sim \mathit{TQG}(0,1,2.9)$ are independent random variables. The exact $95\%$ coverage interval derived via CFA is \linebreak$\operatorname{CI}_{CFA} = [-9.1540 \times 10^{22} , 9.1540\times 10^{22}]$.\label{fig03}}
\end{figure}

This example highlights the importance of having an exact method such as CFA for extreme cases, where MCM may require an impractically large number of realizations to achieve reasonable accuracy. Additionally, the computational cost of using Monte Carlo simulations can be significantly higher compared to the CFA. The accuracy of the Monte Carlo method is highly dependent on the number of realizations, making it computationally infeasible in cases with extremely heavy tailed distributions of the input quantities to obtain a comparably accurate result in a reasonable amount of time. In contrast, the CFA in combination with appropriate numerical algorithms can provide exact results in a numerical sense with relatively low computational cost, making it a preferred choice for many practical applications.

To compare the results, we used MCM with (\ref{eqnXB}), (\ref{eqnXN}) and (\ref{eqnXT}) to derive the result by generating $N=10^5$ realizations of the $X_i$ to obtain $N$ realizations of $Y$. By sorting the realizations of $Y$, we obtained $\operatorname{CI}_{MCM} = [ -1.6820 \times 10^{23} , 1.1835\times 10^{23}]$. These results indicate that $N=10^5$ realizations of the random variable $Y$ are insufficient to ensure accuracy. To obtain a result comparably accurate to CFA, a larger number of realizations is~required. 

For illustration, we increased the number of Monte Carlo simulations one thousand times to $N=10^8$ realizations and we obtained $\operatorname{CI}_{MCM} = [ -9.1671 \times 10^{22} , 9.2231\times 10^{22}]$. Now, this result is comparable to the CFA, but still different and the computation took 67.5~s, whereas it took 0.006 s to obtain the CFA result using the same standard laptop~computer.

Using the CFA in combination with the \href{https://github.com/witkovsky/CharFunTool/blob/master/CF_InvAlgorithms/cf2CDF_GPA.m}{\textbf{cf2CDF\_GPA}} and \href{https://github.com/witkovsky/CharFunTool/blob/master/CF_InvAlgorithms/cf2QF_GPA.m}{\textbf{cf2QF\_GPA}} algorithms provides exact results in a numerical sense. This means that the values obtained are numerically correct and the numerical error can be controlled by appropriately setting the control parameters in the options settings. By using these algorithms, the CDF values and quantiles required to construct a $95\%$ coverage interval can be accurately calculated.
}

\section{Conclusions}
\label{sec6}

{
The Tsallis $q$-Gaussian distribution is a flexible and versatile generalization of the standard Gaussian distribution that can effectively model input quantities in a wide range of applications and measurement models. Its unique properties, such as a finite support for $q < 1$ and an infinite variance for $q \geq 5/3$, make it particularly useful for modeling specific quantities with a known range or extreme values. 

It is important to note that in such situations (with $q \geq 5/3$), the standard uncertainty analysis as specified in GUM may not be applicable, as it implicitly assumes the existence of at least two first moments (mean and variance) of the distributions associated with the input quantities. Therefore, alternative methods are needed to adequately characterize the uncertainty of the output quantities.
}

This paper presents a new and original contribution by providing the explicit form of the characteristic function of $q$-Gaussian random variables and their linear combinations, which can be used in combination with the {Characteristic Function Approach} for efficient and exact uncertainty analysis.

{
The CFA is a powerful tool for deriving the CDF, PDF and QF of the measured quantity, which are essential for specifying measurement uncertainty, in a sense as specified in the Supplements of the GUM and constructing coverage intervals. The CFA offers several advantages, including accuracy, flexibility, efficiency and repeatability.}

To perform these calculations, the MATLAB toolbox \href{https://github.com/witkovsky/CharFunTool}{\textbf{CharFunTool}} is suggested, which includes implemented inversion algorithms such as \href{https://github.com/witkovsky/CharFunTool/blob/master/CF_InvAlgorithms/cf2DistGP.m}{\textbf{cf2DistGP}} or \href{https://github.com/witkovsky/CharFunTool/blob/master/CF_InvAlgorithms/cf2DistGPA.m}{\textbf{cf2DistGPA}}. These algorithms employ the Gil-Pelaez inversion formulae and the adaptive Gauss--Kronrod quadrature rule for numerical integration of the oscillatory integrand function. Additionally, convergence acceleration techniques are employed to compute the limit of the alternating~series.

{We note that finding an explicit expression for the characteristic function of the output quantity and numerically inverting it can be challenging, especially when deriving the joint characteristic function of a multivariate distribution with stochastically dependent input quantities. In such situations, we suggest considering alternative approaches, such as those based on applying the proper Bayesian approach.}

\section*{Acknowledgements}
{The work was supported by the Slovak Research and Development Agency, project APVV-21-0216 and by the Scientific Grant Agency of the Ministry of Education of the Slovak Republic and the Slovak Academy of Sciences, projects VEGA 2/0096/21 and VEGA 2/0023/22.}

 



\bibliographystyle{elsarticle-harv} 


\end{document}